\documentclass[aps,prd,twocolumn,a4paper,superscriptaddress,10pt,nofootinbib]{revtex4-1}
\usepackage{graphicx}
\usepackage{multirow}
\usepackage{latexsym}
\usepackage{amsmath}
\usepackage{amssymb}
\usepackage[colorlinks=true, pdfstartview=FitV, linkcolor=blue, citecolor=red, urlcolor=black, breaklinks=true]{hyperref}
\begin{document}
\newcommand{\be}{\begin{equation}}
\newcommand{\ee}{\end{equation}}
\newcommand{\bq}{\begin{eqnarray}}
\newcommand{\eq}{\end{eqnarray}}

\title{Fundamental physics with ESPRESSO: Constraints on Bekenstein and dark energy models from astrophysical and local probes}
\thanks{Based in part on Guaranteed Time Observations collected at the European Southern Observatory under ESO programme 1102.A-0852 by the ESPRESSO Consortium.}
\author{C.~J.~A.~P.~Martins}
\email{Carlos.Martins@astro.up.pt}
\affiliation{Instituto de Astrof\'isica e Ci\^encias do Espa\c co, CAUP, Universidade do Porto, Rua das Estrelas, 4150-762, Porto, Portugal}
\affiliation{Centro de Astrof\'{\i}sica da Universidade do Porto, Rua das Estrelas, 4150-762 Porto, Portugal}
\author{S.~Cristiani}
\affiliation{INAF -- Osservatorio Astronomico di Trieste, via G. B. Tiepolo 11, I-34143 Trieste, Italy}
\affiliation{Institute for Fundamental Physics of the Universe, Via Beirut 2, I-34151 Miramare, Trieste, Italy}
\author{G.~Cupani}
\affiliation{INAF -- Osservatorio Astronomico di Trieste, via G. B. Tiepolo 11, I-34143 Trieste, Italy}
\affiliation{Institute for Fundamental Physics of the Universe, Via Beirut 2, I-34151 Miramare, Trieste, Italy}
\author{V.~D'Odorico}
\affiliation{INAF -- Osservatorio Astronomico di Trieste, via G. B. Tiepolo 11, I-34143 Trieste, Italy}
\affiliation{Institute for Fundamental Physics of the Universe, Via Beirut 2, I-34151 Miramare, Trieste, Italy}
\author{R.~G\'enova~Santos}
\affiliation{Instituto de Astrof\'{\i}sica de Canarias (IAC), Calle V\'{\i}a L\'actea s/n, E-38205 La Laguna, Tenerife, Spain}
\affiliation{Departamento de Astrof\'{\i}sica, Universidad de La Laguna (ULL), E-38206 La Laguna, Tenerife, Spain}
\author{A.~C.~O.~Leite}
\affiliation{Instituto de Astrof\'isica e Ci\^encias do Espa\c co, CAUP, Universidade do Porto, Rua das Estrelas, 4150-762, Porto, Portugal}
\affiliation{Centro de Astrof\'{\i}sica da Universidade do Porto, Rua das Estrelas, 4150-762 Porto, Portugal}
\author{C.~M.~J.~Marques}
\affiliation{Faculdade de Ci\^encias e Tecnologia, Universidade Nova de Lisboa, 2829-516 Caparica, Portugal}
\affiliation{Centro de Astrof\'{\i}sica da Universidade do Porto, Rua das Estrelas, 4150-762 Porto, Portugal}
\author{D.~Milakovi\'c}
\affiliation{INAF -- Osservatorio Astronomico di Trieste, via G. B. Tiepolo 11, I-34143 Trieste, Italy}
\affiliation{Institute for Fundamental Physics of the Universe, Via Beirut 2, I-34151 Miramare, Trieste, Italy}
\affiliation{Istituto Nazionale di Fisica Nucleare, Sezione di Trieste, Via Bonomea 265, 34136 Trieste, Italy}
\author{P.~Molaro}
\affiliation{INAF -- Osservatorio Astronomico di Trieste, via G. B. Tiepolo 11, I-34143 Trieste, Italy}
\affiliation{Institute for Fundamental Physics of the Universe, Via Beirut 2, I-34151 Miramare, Trieste, Italy}
\author{Michael~T.~Murphy}
\affiliation{Centre for Astrophysics and Supercomputing, Swinburne University of Technology, Hawthorn, Victoria 3122, Australia}
\affiliation{Institute for Fundamental Physics of the Universe, Via Beirut 2, I-34151 Miramare, Trieste, Italy}
\author{N.~J.~Nunes}
\affiliation{Instituto de Astrof\'isica e Ci\^encias do Espa\c{c}o, Faculdade de Ci\^encias da Universidade de Lisboa, Campo Grande, PT1749-016 Lisboa, Portugal}
\affiliation{Departamento de Física da Faculdade de Ciências da Universidade de Lisboa, Edifício C8, 1749-016 Lisboa, Portugal}
\author{Tobias~M.~Schmidt}
\affiliation{Observatoire Astronomique de l'Universit\'e de Gen\`eve, Chemin Pegasi 51, CH-1290 Versoix, Switzerland}
\affiliation{INAF -- Osservatorio Astronomico di Trieste, via G. B. Tiepolo 11, I-34143 Trieste, Italy}
\author{V.~Adibekyan}
\affiliation{Instituto de Astrof\'isica e Ci\^encias do Espa\c co, CAUP, Universidade do Porto, Rua das Estrelas, 4150-762, Porto, Portugal}
\affiliation{Departamento de F\'isica e Astronomia, Faculdade de Ci\^encias, Universidade do Porto, Rua Campo Alegre, 4169-007, Porto, Portugal}
\author{Y.~Alibert}
\affiliation{Physics Institute, University of Bern, Sidlerstrasse 5, 3012 Bern, Switzerland}
\author{Paolo~Di~Marcantonio}
\affiliation{INAF -- Osservatorio Astronomico di Trieste, via G. B. Tiepolo 11, I-34143 Trieste, Italy}
\author{J.~I.~Gonz\'alez~Hern\'andez}
\affiliation{Instituto de Astrof\'{\i}sica de Canarias (IAC), Calle V\'{\i}a L\'actea s/n, E-38205 La Laguna, Tenerife, Spain}
\affiliation{Departamento de Astrof\'{\i}sica, Universidad de La Laguna (ULL), E-38206 La Laguna, Tenerife, Spain}
\author{D.~M\'egevand}
\affiliation{Observatoire Astronomique de l'Universit\'e de Gen\`eve, Chemin Pegasi 51, CH-1290 Versoix, Switzerland}
\author{E.~Palle}
\affiliation{Instituto de Astrof\'{\i}sica de Canarias (IAC), Calle V\'{\i}a L\'actea s/n, E-38205 La Laguna, Tenerife, Spain}
\affiliation{Departamento de Astrof\'{\i}sica, Universidad de La Laguna (ULL), E-38206 La Laguna, Tenerife, Spain}
\author{F.~A.~Pepe}
\affiliation{Observatoire Astronomique de l'Universit\'e de Gen\`eve, Chemin Pegasi 51, CH-1290 Versoix, Switzerland}
\author{N.~C.~Santos}
\affiliation{Instituto de Astrof\'isica e Ci\^encias do Espa\c co, CAUP, Universidade do Porto, Rua das Estrelas, 4150-762, Porto, Portugal}
\affiliation{Departamento de F\'isica e Astronomia, Faculdade de Ci\^encias, Universidade do Porto, Rua Campo Alegre, 4169-007, Porto, Portugal}
\author{S.~G.~Sousa}
\affiliation{Instituto de Astrof\'isica e Ci\^encias do Espa\c co, CAUP, Universidade do Porto, Rua das Estrelas, 4150-762, Porto, Portugal}
\affiliation{Departamento de F\'isica e Astronomia, Faculdade de Ci\^encias, Universidade do Porto, Rua Campo Alegre, 4169-007, Porto, Portugal}
\author{A.~Sozzetti}
\affiliation{INAF -- Osservatorio Astrofisico di Torino, via Osservatorio 20, I-10025 Pino Torinese, Italy}
\author{A.~Su\'arez~Mascare\~no}
\affiliation{Instituto de Astrof\'{\i}sica de Canarias (IAC), Calle V\'{\i}a L\'actea s/n, E-38205 La Laguna, Tenerife, Spain}
\affiliation{Departamento de Astrof\'{\i}sica, Universidad de La Laguna (ULL), E-38206 La Laguna, Tenerife, Spain}
\author{M.~R.~Zapatero~Osorio}
\affiliation{Centro de Astrobiolog\'{\i}a (CSIC-INTA), Crta. Ajalvir km 4, E-28850 Torrej\'on de Ardoz, Madrid, Spain}

\date{19 February 2022}

\begin{abstract}
Dynamical scalar fields in an effective four-dimensional field theory are naturally expected to couple to the rest of the theory's degrees of freedom, unless some new symmetry is postulated to suppress these couplings. In particular, a coupling to the electromagnetic sector will lead to spacetime variations of the fine-structure constant, $\alpha$. Astrophysical tests of the space-time stability of $\alpha$ are therefore a powerful probe of new physics. Here we use ESPRESSO and other contemporary measurements of $\alpha$, together with background cosmology data, local laboratory atomic clock and Weak Equivalence Principle measurements, to place stringent constraints on the simplest examples of the two broad classes of varying $\alpha$ models: Bekenstein models and quintessence-type dark energy models, both of which are parametric extensions of the canonical $\Lambda$CDM model. In both cases, previously reported constraints are improved by more than a factor of ten. This improvement is largely due to the very strong local constraints, but astrophysical measurements can help to break degeneracies between cosmology and fundamental physics parameters.
\end{abstract}
\maketitle

%%%%%%%%%%%%%%%%%%%%%%%%%%%%%%%%%%%%%%%%%%%%%%%%%%%%%%%%%%%%%%%%%%%%%%%%%%
\section{\label{intr}Introduction}

Dynamical scalar fields are ubiquitous in some commonly considered types of fundamental physics theory, and they will naturally couple to the rest of the theory's degrees of freedom. For example, these couplings unavoidably exist in string theory \cite{Taylor}. While several dimensionless fundamental couplings can be theoretically expected to vary, and such possible variations can be constrained both by local experiments and by high-resolution astrophysical spectroscopy, in this work we focus on the coupling with the electromagnetic sector, which would lead to three inter-related consequences: a time (redshift) dependence of the fine-structure constant, $\alpha$, a violation of the Einstein Equivalence Principle \citep{Carroll,Dvali,Chiba,Dilaton}, and a fifth force of nature---see \cite{Damour} and references therein. In this work we address the first two. A detection of such effects would be revolutionary, but as we show in the present work even improved null results are extremely useful.

In the last two decades high-resolution astrophysical spectroscopy tests of the stability of $\alpha$, done along the line of sight of bright quasars, are a source of much interest and also some controversy, summarized in recent reviews \cite{Uzan,ROPP}. The new high-resolution spectrograph at the VLT, ESPRESSO \cite{Pepe}, was specifically designed with the goal of resolving this controversy \cite{Molaro}, \textit{inter alia} by drastically reducing wavelength calibration errors by using a laser frequency comb \cite{Schmidt}. The ESPRESSO Consortium's Guaranteed Time Observations include a program of measurements of $\alpha$, the first of which (along the line of sight of HE\,0515$-$4414, one of the brightest quasars in the southern sky) has recently been published \cite{Murphy}. Here we report on the impact of this measurement, together with other contemporary measurements of $\alpha$, background cosmology data and local laboratory tests, on models of fundamental cosmology.

Phenomenologically, realistic models for varying couplings can be divided into two classes \cite{ROPP}. The first, dubbed Class I, contains models where the degree of freedom responsible for varying $\alpha$ (typically a scalar field) also provides the dark energy. These are arguably the minimal models, in the operational sense that a single new dynamical degree of freedom---in other words, a single extension of the standard model---accounts for both. Conversely, in Class II models the field that provides the varying $\alpha$ does not provide the dark energy (or at least does not provide all of it).

In what follows, after summarizing the datasets that we use, we present updated constraints on the simplest representative models in each of the two classes, respectively Bekenstein models (the simplest class of Class II models) and quintessence-type dark energy models (the most studied example of a Class I model). Both of these are parametric extensions of the canonical $\Lambda$CDM model (in the sense that that latter model is recovered for specific choices of the model parameters), and therefore our analysis constraints the level of deviations from $\Lambda$CDM allowed by these datasets and shows that these must be very small.

%%%%%%%%%%%%%%%%%%%%%%%%%%%%%%%%%%%%%%%%%%%%%%%%%%%%%%%%%%%%%%%%%%%%%%%%%%
\section{\label{data}Relevant datasets}

In order to optimally constrain the models (and reduce degeneracies between model parameters) the astrophysical spectroscopy measurements should be combined with external datasets, and in the present work we also include cosmological and local experiment data. We now describe our assumptions for each of these.

Since the spectroscopic measurements of $\alpha$ along the line of sight of bright quasars are akin to background cosmology observations, in choosing our cosmological datasets we also restrict ourselves to background cosmology data. Using cosmological data from clustering observations would require further assumptions on the impact of possible $\alpha$ variations therein, and we leave this for separate work. Bearing this in mind we will use two separate low-redshift background cosmology datasets, both of which have been extensively used in the literature in recent years. The first subset is the Pantheon Type Ia supernova compilation \cite{Riess}. This is a 1048 supernova dataset, containing measurements in the range $0<z<2.3$, further compressed into 6 correlated measurements of $E^{-1}(z)$ (where $E(z)=H(z)/H_0$ is the dimensionless Hubble parameter) in the redshift range $0.07<z<1.5$. This provides an effectively identical characterization of dark energy as the full supernova sample, thus making it an efficient compression of the raw data. The second subset is a compilation of 38 Hubble parameter measurements \cite{Farooq}. In our analysis the two subsets will always be used together, making what we will refer to as the \textit{Cosmology} dataset. We note that the Hubble constant was analytically marginalized in the analysis, following the procedure in \cite{Anagnostopoulos}, so the much debated Hubble tension does no impact our results.

Constraints on $\alpha$ at a given redshift are usually expressed relative to the present-day laboratory value $\alpha_0$, specifically via
$(\Delta\alpha/\alpha)(z)\equiv(\alpha(z)-\alpha_0)/\alpha_0$, with competitive measurements being at the parts per million (ppm) level. Direct high-resolution spectroscopy measurements of $\alpha$ are done (mainly at optical wavelengths) in low-density absorption clouds along the line of sight of bright quasars (QSOs). We emphasize that these are direct and model-independent measurements. In what follows we will separately consider two subsets of these measurements.

The first $\alpha$ subset is the dataset of Webb \textit{et al.} \cite{Webb}, which we henceforth refer to as the \textit{Archival} dataset. This is a dataset of 293 measurements from VLT-UVES and Keck-HIRES. The data were originally taken for other purposes and subsequently reanalysed by the authors for the purpose of measuring $\alpha$. This is relevant because $\alpha$ measurements require particularly careful wavelength calibration procedures, with rely on additional data, coeval with the quasar observations. Such additional data is not ordinarily taken for observations which do not have the stringent requirements for $\alpha$ tests, and cannot be obtained \textit{a posteriori}. Moreover, unknown at the time of the original analysis, the spectrographs providing these data are now known to suffer from significant wavelength distortions \cite{Whitmore}. Such limitations may be partially mitigated \cite{Dumont}, but cannot be fully eliminated.

The second $\alpha$ subset, which we call the \textit{Dedicated} dataset, contains 30 measurements obtained for the purpose of constraining $\alpha$, where ancillary data enabled a more robust wavelength calibration procedure, or using more modern spectrographs that do not suffer from the limitations of VLT-UVES or Keck-HIRES. In addition to the measurements listed\footnote{This is a compilation of measurements, from several authors, published between 2013 and 2017 \cite{Agafonova11,Molaro13,Songaila14,Evans14,Murphy16,Bainbridge16,Kotus17}.} in Table 1 of \cite{ROPP} this includes more recent ones from the Subaru telescope \cite{Cooksey}, the HARPS spectrograph \cite{Milakovic}, and two ESPRESSO measurements: our own recently published measurement \cite{Murphy} 
\be
\left(\frac{\Delta\alpha}{\alpha}\right)_{z=1.15}=1.31\pm1.36\,\text{ppm}\,
\ee
(where statistical and systematic uncertainties have been added in quadrature) and an earlier, though much less precise one from Science Verification \cite{Welsh}.

The main reason for treating the Archival and Dedicated datasets separately is that they are discrepant. A simple way to see this is to assume that there is a unique astrophysical value of $\Delta\alpha/\alpha$, which we estimate by taking the weighted mean of all the values in each dataset. In that case we find $\Delta\alpha/\alpha=-2.16\pm0.85$ ppm and $\Delta\alpha/\alpha=-0.23\pm0.56$ ppm respectively for the Archival and Dedicated datasets. Additional comparisons of the two datasets can be found in \cite{ROPP,Cosmography}.

Finally, our \textit{Local} dataset comprises three different constraints. The first is the geophysical constraint from the Oklo natural nuclear reactor\footnote{Note that this constraint is model-dependent, since it only holds under the assumption that $\alpha$ is the only varying coupling.} \cite{OKLO}, at an effective redshift $z_{Oklo}=0.14$. The second comes from laboratory tests comparing atomic clocks based on transitions with different sensitivities to $\alpha$, which lead to a constraint on its current drift rate \cite{Lange}
\be\label{clocks}
\left(\frac{\dot\alpha}{\alpha}\right)_0=(1.0\pm1.1)\times10^{-18}\,\text{yr}^{-1}\,.
\ee
We can also express this as a dimensionless number by dividing it by the Hubble constant, for which we use $H_0=70$  km/s/Mpc. We then find
\be\label{clocksh0}
\frac{1}{H_0}\left(\frac{\dot\alpha}{\alpha}\right)_0=0.014\pm0.015\,\text{ppm}\,,
\ee
highlighting the fact that this is the most stringent individual constraint. Nevertheless, note that if Eq. (\ref{clocks}) is used directly it is model-independent, while if one uses Eq. (\ref{clocksh0}) there is some implicit model dependence. Finally, we use the recent MICROSCOPE bound on the E\"otv\"os parameter, $\eta$, reported in \cite{Touboul},
\be\label{micro}
\eta=(-0.1\pm1.3)\times10^{-14}\,;
\ee
the two test masses are platinum and titanium alloys\footnote{Specifically the first is made of 90\% by mass of platinum and 10\% of rhodium, while the second  is made of 90\% of titanium, 6\% of aluminium and 4\% of vanadium.}. This bound constrains the model's coupling to the electromagnetic sector $\zeta$, to be defined in the following sections, with the relation between the two being model-dependent.

In passing, we also note that additional measurements of $\alpha$ can be obtained at higher redshifts---again this is further discussed in recent reviews \cite{Uzan,ROPP}. The cosmic microwave background provides a constraint at an effective redshift $z_{\rm CMB}\sim1100$, while big bang nucleosynthesis provides a constraint at an effective redshift $z_{\rm BBN}\sim4\times10^8$. The most recent such constraints are \cite{Hart} for the former and \cite{Deal} for the latter. However, these constraints are unavoidably model-dependent (unlike the QSO measurements) and therefore will not be included in our analysis. In practical terms this is a moot point for the CMB case because the constraint is extremely weak (with an uncertainty at the parts per thousand level, as opposed to parts per million) and therefore it would have no statistical weight in our analysis.

%%%%%%%%%%%%%%%%%%%%%%%%%%%%%%%%%%%%%%%%%%%%%%%%%%%%%%%%%%%%%%%%%%%%%%%%%%
\section{\label{beken}Constraints on Bekenstein models}

Arguably the simplest class of phenomenological models for varying $\alpha$ is the one first suggested by Bekenstein \cite{Bekenstein,SBM} where, by construction, the dynamical scalar field $\psi$ responsible for this variation has a negligible effect on the cosmological dynamics, making it a Class II model. These models have a single phenomenological dimensionless parameter, denoted $\zeta$, the coupling of the dynamical scalar degree of freedom to the electromagnetic sector. Also by construction, these models assume that $\alpha$ is the only fundamental coupling that varies, while other parameters, e.g. particle masses, do not. 

Assuming a flat, homogeneous and isotropic cosmology one obtains the following Friedmann and scalar field equations \cite{SBM,LeiteBek}
\be
H^2=\frac{8\pi G}{3}\left[\rho_m(1+\zeta e^{-2\psi})+\rho_r e^{-2\psi}+\rho_\Lambda+\frac{1}{2}{\dot\psi}^2  \right]\,
\ee
\be
{\ddot\psi}+3H{\dot\psi}=-2\zeta G\rho_m e^{-2\psi} \,,
\ee
with the dots denoting derivatives with respect to physical time, and the $\rho_i$ respectively denoting the matter, radiation and dark energy densities. The model needs a dark energy component to match cosmological observations, which for simplicity is assumed to be a cosmological constant. Therefore setting $\zeta=0$ one recovers the canonical $\Lambda$CDM model. The value of $\alpha$ is related to the scalar field $\psi$ via $\alpha/\alpha_0= e^{2(\psi-\psi_0)}$, and without loss of generality we can re-define the field such that at the present day $\psi_0=0$. In practice it is more convenient to write the scalar field equation as a function of redshift
\be\label{dyna}
\psi''+\left(\frac{d\ln{E(z)}}{dz}-\frac{2}{1+z}\right)\psi'=-\frac{3\zeta \Omega_m}{4\pi}\frac{(1+z)}{E^2(z)}  e^{-2\psi}\,;
\ee
here the primes denote derivatives with respect to redshift. Moreover, in this type of model the relation between $\eta$ and the coupling parameter is \cite{SBM}
\be\label{etab}
\eta\sim3\times 10^{-9}\zeta\,.
\ee

%%%%%%%%%%%%%%%%%%%%%%%%%%%%%%%%%
\begin{figure*}
\centering
  \includegraphics[width=1.0\columnwidth]{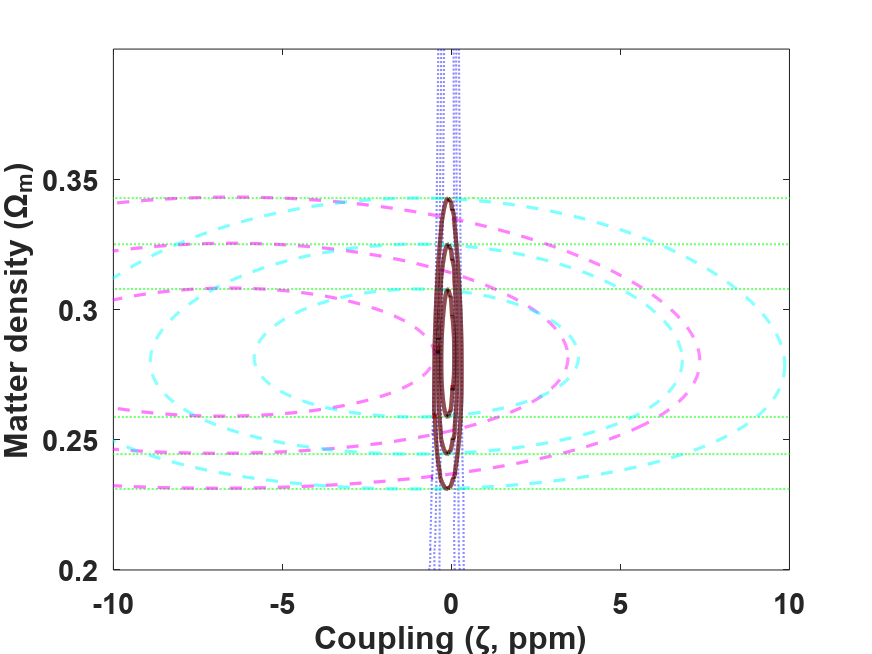}
  \includegraphics[width=1.0\columnwidth]{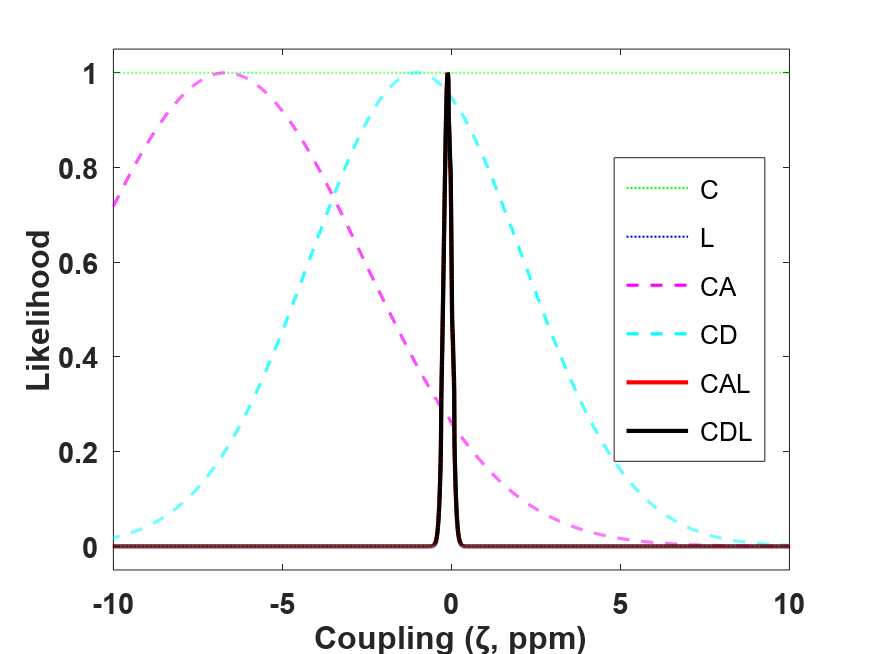}
  \caption{Constraints on the Bekenstein model. The left panel shows one, two and three sigma confidence level contours on the $\Omega_m$--$\zeta$ plane, and the right panel shows the posterior likelihoods for $\zeta$, with $\Omega_m$ marginalized. The colors and line styles denote the following data combinations: Cosmology only (thin dotted green), Local only (thin dotted blue), Cosmology+Archival (dashed magenta), Cosmology+Dedicated (dashed cyan), Cosmology+Archival+Local (thick solid red), Cosmology+Dedicated+Local (thick solid black). Note that the Local only case (in the right panel) and the Cosmology+Archival+Local case (in both panels) are not visible in the plot since they overlap with the Cosmology+Dedicated+Local case.}
  \label{fig1}
\end{figure*}
%%%%%%%%%%%%%%%%%%%%%%%%%%%%%%%%%
\begin{table*}
\begin{tabular}{| c | c | c | c |}
\hline
Datasets & Figure 1 Panels & $\Omega_m$ & $\zeta$ (ppm) \\
\hline
Cosmology only & Thin dotted green & $0.28\pm0.02$ & Unconstrained \\
Local only & Thin dotted blue & Unconstrained & $-0.10\pm0.11$ \\
\hline
Cosmology+Archival & Dashed magenta & $0.28\pm0.02$ & $-6.7\pm4.1$ \\
Cosmology+Dedicated & Dashed cyan & $0.28\pm0.02$ & $-1.0\pm3.2$ \\
\hline
Cosmology+Archival+Local & Thick solid red & $0.28\pm0.02$ & $-0.11\pm0.12$ \\
Cosmology+Dedicated+Local & Thick solid black & $0.28\pm0.02$ & $-0.11\pm0.12$ \\
\hline
\end{tabular}
\caption{Constraints on the parameters of the Bekenstein model, for various combinations of datasets. The colors and line styles refer to the panels of Fig. \protect\ref{fig1}.\label{table1}}
\end{table*}
%%%%%%%%%%%%%%%%%%%%%%%%%%%%%%%%%

Figure \ref{fig1} and Table \ref{table1} show our constraints for various data combinations. The cosmological data only constrains the matter density, while the local data only constrains the coupling, since that will affect the field speed today. Combining the cosmology and spectroscopic data one can constrain both parameters, without a significant correlation between them. The Archival data has a small preference for a negative coupling, while the Dedicated one is consistent with a null coupling. Note that both of these coupling constraints are at the level of few ppm, comparable to the constraint from the E\"otv\"os parameter, which one can obtain by comparing Eq. (\ref{micro}) with Eq. (\ref{etab}).

In any case, in the full dataset the local data dominates the constraints, to the extent that they become identical for the Archival and Dedicated data, and there is no indication of a non-zero coupling of this kind. The one-sigma uncertainty is at the 0.12 ppm level, which improves on earlier constraints in \cite{LeiteBek} and \cite{ROPP} by factors of 14 and 12 respectively. Thus, for these models, where the cosmological and particle physics parameters are not significantly correlated, contemporary astrophysical measurements of $\alpha$ do not play a significant role.

%%%%%%%%%%%%%%%%%%%%%%%%%%%%%%%%%%%%%%%%%%%%%%%%%%%%%%%%%%%%%%%%%%%%%%%%%%
\section{\label{darken}Constraints on dark energy models}

Class I models assume that the same degree of freedom provides both the dark energy and the varying $\alpha$. One consequence of this is that the cosmological evolution of the latter is parametrically determined. Specifically, assuming a canonical scalar field (with a $w_\phi(z)\ge-1$), one finds that \cite{Erminia1}
\begin{equation} \label{eq:dalfa}
\frac{\Delta\alpha}{\alpha}(z) =\zeta \int_0^{z}\sqrt{3\Omega_\phi(z')\left(1+w_\phi(z')\right)}\frac{dz'}{1+z'}\,,
\end{equation}
where $w_\phi(z)$ is the dark energy equation of state and $\Omega_\phi (z) \equiv \rho_\phi(z)/\rho_{\rm tot}(z)\simeq \rho_\phi(z)(\rho_\phi(z)+\rho_m(z))$ is the fraction of the dark energy density, where in the last step we have neglected the contribution from radiation, since we are interested in low redshifts. For phantom fields (with $w_\phi(z)<-1$) one has instead \cite{Phantom}
\begin{equation} \label{eq:dalfa2}
\frac{\Delta\alpha}{\alpha}(z) =-\zeta \int_0^{z}\sqrt{3\Omega_\phi(z')\left|1+w_\phi(z')\right|}\frac{dz'}{1+z'}\,;
\end{equation}
the change of sign stems from the fact that one expects phantom fields to roll up the potential rather than down. From this we find the present-day drift rate of $\alpha$,
\begin{equation} \label{clocks2}
\frac{1}{H_0}\frac{\dot\alpha}{\alpha} =\, \mp \, \zeta\sqrt{3\Omega_{\phi0}|1+w_0|}\,,
\end{equation}
where $w_0$ is the present-day dark energy equation of state, with the minus and plus signs respectively corresponding to the canonical and phantom field cases. This shows that there will be a degeneracy between the coupling $\zeta$ and the parameters describing the dynamics of dark energy, which naturally did not occur for the models in the previous section. On the other hand, there is again no significant correlation of $\zeta$ with the matter density, and for this reason we assume a fixed value of $\Omega_m=0.3$ in this section. For this class of models $\eta$ and the dimensionless coupling $\zeta$ are simply related by \cite{Dvali,Chiba,Damour}
\begin{equation} \label{eq:eotvos}
\eta \approx 10^{-3}\zeta^2\,.
\end{equation}

Specifically, we consider the Chevallier-Polarski-Linder (CPL) parametrization for the dark energy equation of state \cite{CPL1,CPL2}
\begin{equation} \label{cpl}
w_{\rm CPL}(z)=w_0+w_a \frac{z}{1+z}\,,
\end{equation}
where $w_0$ is its present value and $w_a$ is the coefficient of the time-dependent term. This is a phenomenological parametrization, assumed to be representative of dynamical scalar fields, and allowing for canonical and phantom equations of state. In addition to its frequent use for dark energy studies, it is also often used for varying $\alpha$ studies, being the prototypical example of a Class I model. For example, it is the fiducial model adopted in recent forecasts of cosmological constraints on dark energy from the combination of QSO measurements of $\alpha$ with cosmological data from the Euclid satellite \cite{Euclid}. Assuming a flat universe the Friedmann equation can be written
\begin{equation}
E^2(z)=\Omega_m(1+z)^3+(1-\Omega_m)(1+z)^{3(1+w_0+w_a)}e^{-3w_az/(1+z)}\,;
\end{equation}
again the canonical $\Lambda$CDM case is recovered for $w_0=-1$ and $w_a=0$, together with $\zeta=0$.

In this case, and in addition to varying $\alpha$, other fundamental couplings could vary. In what follows we only use $\alpha$ measurements to constrain the model because so far in the ESPRESSO consortium we have only published measurements of $\alpha$. Including QSO or atomic clock constraints on further parameters such as the proton-to-electron mass ratio ( again see \cite{Uzan,ROPP} for reviews on these) would further improve the constraints which we report in this section, but that is left for future work. In this sense, our constraints on this model can be seen as conservative constraints.

%%%%%%%%%%%%%%%%%%%%%%%%%%%%%%%%%
\begin{figure*}
\centering
  \includegraphics[width=1.0\columnwidth]{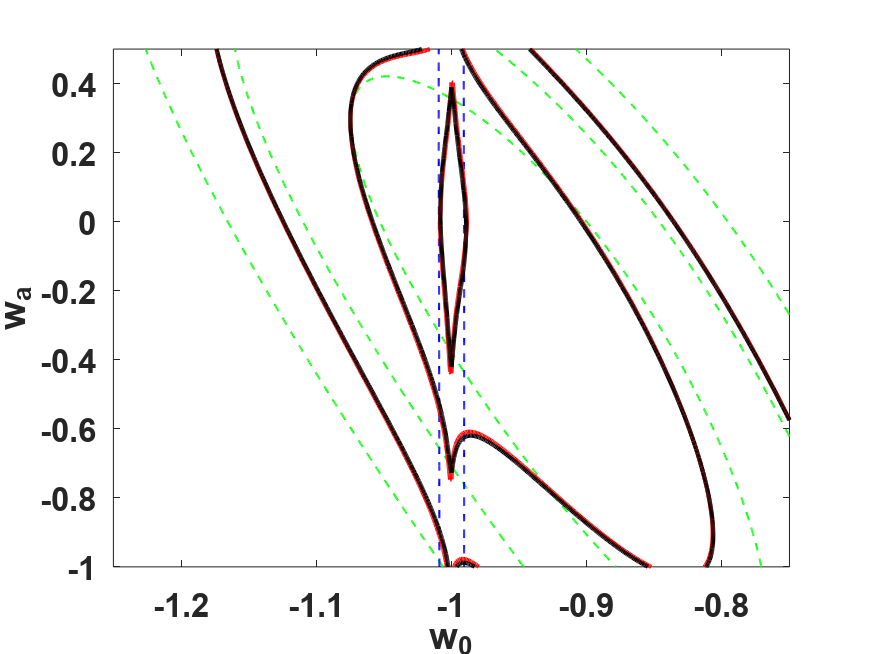}
  \includegraphics[width=1.0\columnwidth]{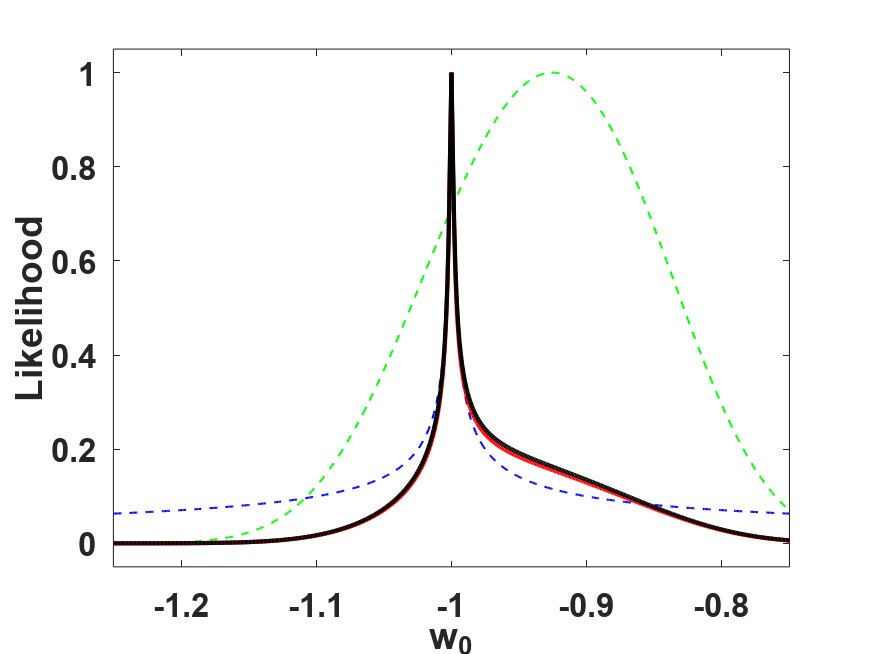}
  \includegraphics[width=1.0\columnwidth]{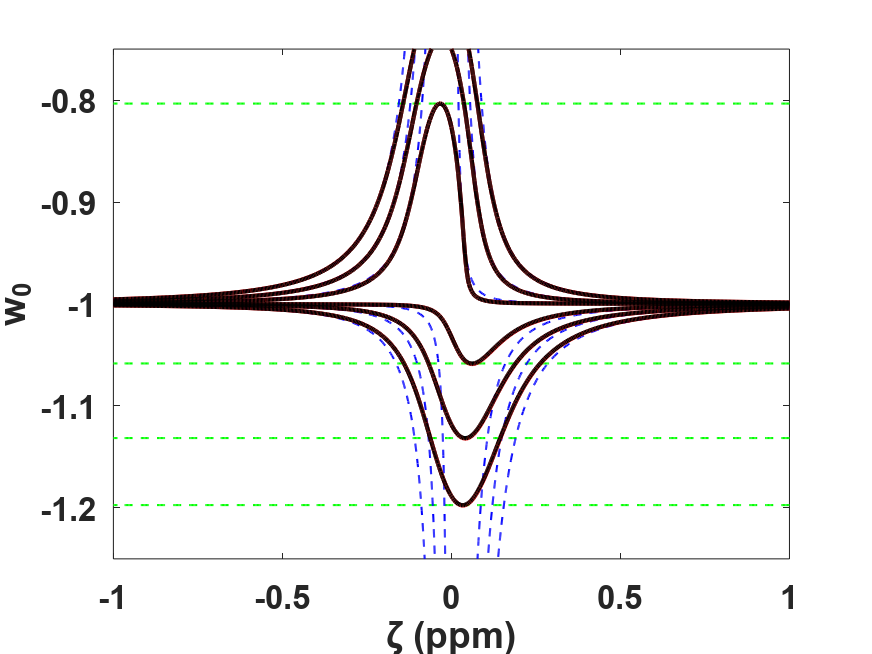}
  \includegraphics[width=1.0\columnwidth]{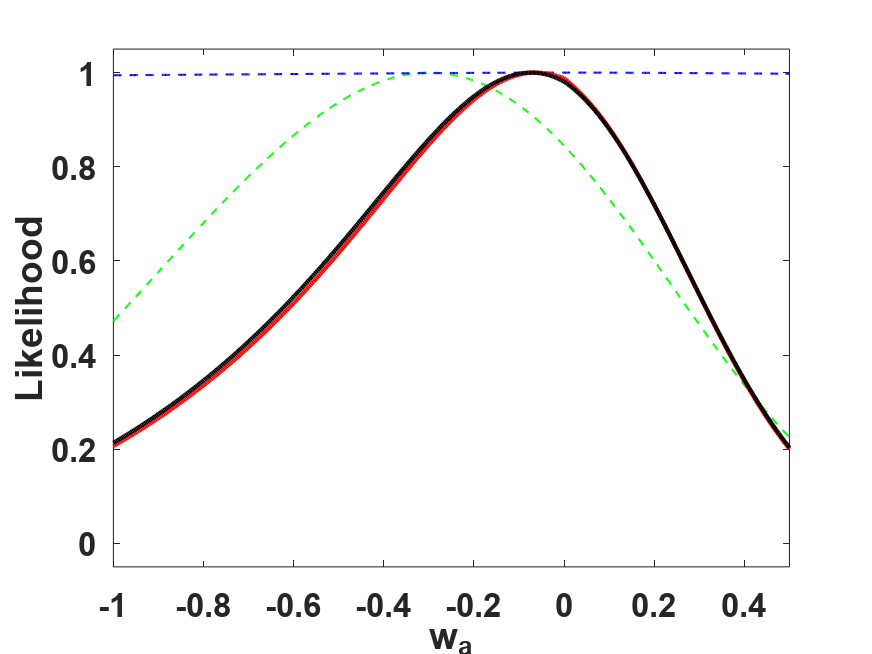}
  \includegraphics[width=1.0\columnwidth]{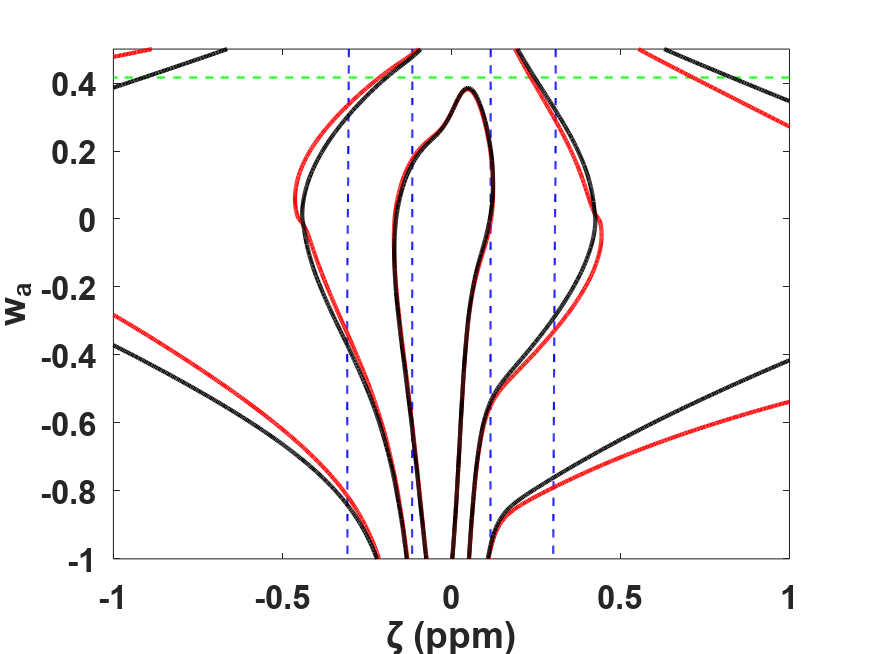}
  \includegraphics[width=1.0\columnwidth]{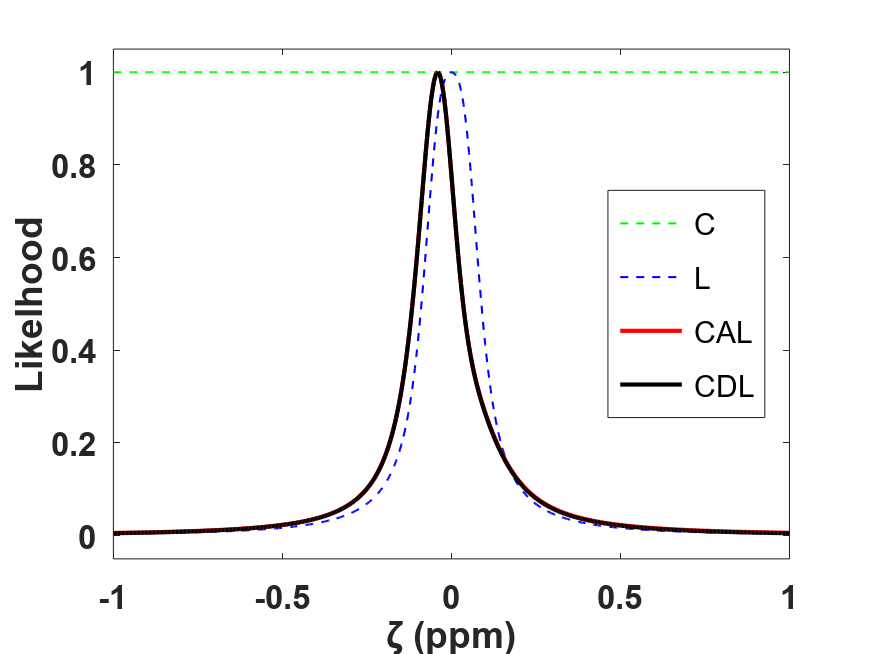}
  \caption{Constraints on the CPL parametrization. The left panels show one, two and three sigma confidence level contours on the relevant two-dimensional planes, and the right panels show the posterior likelihoods for each parameter with the others marginalized. The colors and line styles denote the following data combinations: Cosmology only (thin dashed green), Local only (thin dashed blue), Cosmology+Archival+Local (thick solid red), Cosmology+Dedicated+Local (thick solid black). Note that the constraints for the Cosmology+Archival+Local and the Cosmology+Dedicated+Local are very similar so in most panels the curves for the former are not easily visible since they overlap with those of the latter.}
  \label{fig2}
\end{figure*}
%%%%%%%%%%%%%%%%%%%%%%%%%%%%%%%%%
\begin{table*}
\begin{tabular}{| c | c | c | c | c |}
\hline
Datasets & Figure 2 Panels & $w_0$ & $w_a$ & $\zeta$ (ppm) \\
\hline
Cosmology only & Thin dashed green & $-0.93\pm0.09$ & $-0.30^{+0.50}_{-0.57}$ & Unconstrained \\
Local only & Thin dashed blue & $-1.00\pm0.12$ & Unconstrained & $0.00\pm0.09$ \\
\hline
Cosmology+Archival+Local & Thick solid red & $-1.00^{+0.09}_{-0.04}$ & $-0.06^{+0.32}_{-0.45}$ & $-0.04^{+0.14}_{-0.11}$ \\
Cosmology+Dedicated+Local & Thick solid black & $-1.00^{+0.09}_{-0.04}$ & $-0.07^{+0.33}_{-0.45}$ & $-0.04^{+0.14}_{-0.11}$ \\
\hline
\end{tabular}
\caption{Constraints on the CPL parametrization. The colors and line styles refer to the panels of Fig. \protect\ref{fig2}.\label{table2}}
\end{table*}
%%%%%%%%%%%%%%%%%%%%%%%%%%%%%%%%%

Figure \ref{fig2} and Table \ref{table2} show the constraints for this case. Now we have three relevant parameters, and in addition to the wider parameter space the three parameters are correlated, i.e. there are significant degeneracies between them. The cosmological data can constrain the two dark energy parameters but not $\zeta$. As can be seen from Eq. (\ref{clocks2}), atomic clocks only constrain a combination of $w_0$ and $\zeta$, with this degeneracy being broken by the E\"otv\"os parameter constraint. And although the Oklo constraint is nominally sensitive to all three parameters, its very low redshift lever arm means that the combination of the three local measurements leaves $w_a$ unconstrained. Therefore, to simplify the discussion (and the legibility of Fig. \ref{fig2}), we compare only the cases of cosmology data, local data, and the full dataset (Cosmology plus QSO plus Local data), still separating the archival and dedicated QSO measurements.

Here it is still the case that the local data dominates the overall constraints, but nevertheless the addition of the spectroscopic data does have an impact in skewing the relative preferences between the canonical ($w(z)\ge-1$) and phantom ($w(z)<-1$) regions of the parameter space. While the posterior likelihood for $w_0$ from the local data is essentially symmetric around $w_0=-1$ and the cosmology data prefers a somewhat negative $w_a$, in the full data the phantom side of the $w_0$ likelihood is suppressed and the peak likelihood of $w_a$ shifts closer to zero. Admittedly the statistical significance of these differences is not high, but they do highlight the importance of having data spanning a large redshift lever arm in order to constrain these models, and astrophysical measurements of $\alpha$ can therefore play an important role here. In other words, since the putative scalar field is constrained to be evolving slowly (if at all) with redshift, it is important to map its behaviour over a redshift range that is as wide as possible.

Note that the constraint on the coupling $\zeta$ obtained from the local data becomes weaker when the rest of the data is added. This is to be expected, and has previously been discussed in the literature \cite{Euclid1,Euclid2}: given the degeneracy between the dark energy equation of state parameters and the coupling, both of which need to be non-trivial to enable an $\alpha$ variation---cf. Eq. (\ref{eq:dalfa})---if all the data is consistent with the standard model then improving constraints on one sector weakens the constraints on the other sector. Here the constraints on $w_0$ and $w_a$ are improved by the data combination, while that on the coupling is slightly weakened. In any case, for the overall constraint on $\zeta$, earlier constraints in \cite{Pinho,ROPP} are also improved by a factor of 12.

\section{\label{conc}Conclusions}

Dynamical scalar fields in an effective four-dimensional field theory are naturally expected to couple to the rest of the theory, unless a still unknown symmetry is postulated to suppress these couplings. We have used a combination of cosmological, spectroscopic, and local laboratory and low Earth orbit tests to place the most stringent tests on such couplings to the electromagnetic sector, in the context of the simplest examples of the two classes of such models: Bekenstein models and quintessence-type dark energy models. In both models considered we improved previously reported constraints by more than a factor of ten, showing that such couplings can be no larger than parts per million level.

These constraints are dominated by the local data, specifically by the atomic clocks and MICROSCOPE bounds (with Oklo playing a minor role). Given that both of these constraints are expected to be further improved in the near future, one may wonder about the role of astrophysical tests of the stability of $\alpha$, as carried out by ESPRESSO and its forthcoming successor, ANDES\footnote{This is the ArmazoNes high Dispersion Echelle Spectrograph, the recently given name of the high-resolution optical and infrared spectrograph formerly known as ELT-HIRES.}. Apart from the conceptual importance of an independent test of the Weak Equivalence Principle and Local Position Invariance (complementing those done in local laboratories or in the solar system), our analysis indicates that the broad redshift range they provide is important in breaking degeneracies between the dark cosmology and fundamental physics sectors in various classes of models where both of these sectors impact $\alpha$ variations.

As an example, it has been previously suggested \cite{VilasBoas} that sufficiently sensitive $\alpha$ data can distinguish between freezing and thawing models of dark energy, and our analysis of the CPL model is consistent with those findings---for example, one can notice in the bottom left panel of Fig. \ref{fig2} that the Archival and Dedicated datasets lead to small differences in the two-dimensional $\zeta$--$w_a$ plane constraints. Another example would be particle physics or string theory inspired models where the scalar field has different couplings to the baryonic and dark sectors: in such a scenario, local tests will only constrain baryonic sector couplings, leaving dark sector couplings to be constrained by astrophysical and cosmological data.

These are extremely strong constraints, bearing in mind that in most beyond the standard model paradigms, including string theory, these couplings, if they are nonzero, would naively be expected to be of order unity \cite{Taylor,Dilaton,Damour}. At this point it worth to consider three different cosmological settings where dynamical scalar fields play a role. In inflation the field needs to be dynamical (so inflation can end) but it must also slow-roll, at least in the simplest models thereof (cf. the canonical slow-roll inflation conditions). On the other hand, for dark energy and varying $\alpha$ there is currently no evidence of rolling, but there are very stringent constraints on the speed of the putative scalar field, with the constraints for varying $\alpha$ being significantly stronger than those for uncoupled dark energy. Certainly, such couplings of order unity are completely ruled out. The theoretical implications of which remain to be explored.

Finally, the two models which we have constrained are parametric extensions of the $\Lambda$CDM model, and our results constrain such deviations to be very small. If, as many cosmologists expect, the $\Lambda$CDM model is only a simple approximation to a still unknown underlying paradigm, then our results provide further evidence for the point that at a purely phenomenological level $\Lambda$CDM is a remarkably good approximation, and at least at low redshifts any viable extended model must be observationally very similar to it.

\begin{acknowledgments}
This work was done in the context of the CosmoESPRESSO project, financed by FEDER---Fundo Europeu de Desenvolvimento Regional funds through the COMPETE 2020---Operational Programme for Competitiveness and Internationalisation (POCI), and by Portuguese funds through FCT - Funda\c c\~ao para a Ci\^encia e a Tecnologia under project POCI-01-0145-FEDER-028987 and PTDC/FIS-AST/28987/2017 (CJM), with additional support from projects PTDC/FIS-AST/0054/2021 (NJN), PTDC/FIS-AST/28953/2017, POCI-01-0145-FEDER-028953, PTDC/FIS-AST/32113/2017 and POCI-01-0145-FEDER-032113 (NCS, SGS), Investigador FCT contracts IF/00650/2015/CP1273/CT0001 (VA) and CEECIND/00826/2018 (SGS) and UID/FIS/04434/2019, UIDB/04434/2020 and UIDP/04434/2020. The INAF authors (SC, GC, VDO, DM, PM, PDM, AS) acknowledge financial support of the Italian Ministry of Education, University, and Research with PRIN 201278X4FL and the Progetti Premiali funding scheme. MTM acknowledges the support of the Australian Research Council through Future Fellowship grant FT180100194. TMS acknowledges the support from the SNF synergia grant CRSII5-193689 (BLUVES), with additional support of the National Centre of Competence in Research PlanetS supported by the Swiss National Science Foundation. FAP would like to acknowledge the Swiss National Science Foundation (SNSF) for supporting research with ESPRESSO through the SNSF grants nr. 140649, 152721, 166227 and 184618; the ESPRESSO Instrument Project was partially funded through SNSF's FLARE Programme for large infrastructures. ASM acknowledges financial support from the Spanish Ministry of Science and Innovation (MICINN) under 2018 Juan de la Cierva program IJC2018-035229-I, from the MICINN project PID2020-117493GB-I00 and from the Government of the Canary Islands project ProID2020010129. MRZO acknowledges funding under project PID2019-109522GB-C51 of the Spanish Ministerio de Ciencia e Investigaci\'on.

The authors acknowledge the ESPRESSO project team for its effort and dedication in building the ESPRESSO instrument.
\end{acknowledgments}

\bibliography{espresso}
\end{document}